\documentclass{article}

 \usepackage[final]{tackling_climate_workshop_style}
\usepackage{natbib}
\usepackage{float}
\usepackage{graphicx}
\usepackage{xurl}
\usepackage{xfp}
\usepackage{amsmath}
\usepackage[utf8]{inputenc} 
\usepackage[T1]{fontenc}    
\usepackage{hyperref}  
\usepackage{cleveref}
\usepackage{url}            
\usepackage{booktabs}       
\usepackage{amsfonts}       
\usepackage{nicefrac}       
\usepackage{microtype}      

\newcommand{\FASSTSSTmean}{291.30}
\newcommand{\FASSTSSTstd}{0.22}
\newcommand{\CESMSSTmean}{291.25}
\newcommand{\CESMSSTstd}{0.10}

\newcommand{\FASSTNinoStd}{0.62}

\newcommand{\CESMNinoStd}{1.02}

\newcommand{\SpinupYear}{31}
\newcommand{\SpinupThreshold}{0.02}
\newcommand{\SpinupWindow}{15}
\newcommand{\TotalSimulationLength}{140}

\title{FASST: A Compact Stateless Atmosphere Model for Long-Term Ocean Coupling}

\author{%
  Pratham Aggarwal \\
  University of California San Diego\\
  \texttt{praggarwal@ucsd.edu} \\
  \And
  Duncan Watson-Parris \\
  University of California San Diego \\
  \texttt{dwatsonparris@ucsd.edu} \\
  \AND
  William E. Chapman \\
  University of Colorado Boulder \\
  \texttt{wchapman@colorado.edu} \\
}

\begin{document}

\maketitle

\begin{abstract}
Coupled climate simulations are computationally expensive, largely because of the atmosphere. We introduce FASST, a coupled climate model that retains the prognostic POP2 ocean while replacing the atmosphere and sea ice with lightweight U-Net emulators. The atmospheric emulator is diagnostic, predicting surface fluxes from the evolving ocean--ice state every $6$ simulated hours, while the sea-ice emulator advances the ice state monthly. For $\TotalSimulationLength$ years of unforced integration, FASST maintains a stable global-mean climate without SST restoring, with post-transient global-mean SST of $\FASSTSSTmean \pm \FASSTSSTstd~\mathrm{K}$ compared with $\CESMSSTmean \pm \CESMSSTstd~\mathrm{K}$ in CESM$2$ piControl. FASST also captures interannual Ni\~no-$3.4$ variability, although with weaker amplitude than CESM ($\FASSTNinoStd~\mathrm{K}$ versus $\CESMNinoStd~\mathrm{K}$). These results demonstrate that a stateless learned atmosphere can sustain stable, free-running coupled climate simulations over century-scale integrations.
\end{abstract}
\section{Introduction}

Understanding how the ocean absorbs heat and adjusts to climate change requires coupled ocean--atmosphere models that can be integrated for centuries. However, the atmosphere is a major computational bottleneck. In the CAMulator benchmark manuscript, the Community Atmosphere Model version 6 (a full physics fortran based simulation) achieves only $14$ simulated years per day on ten CPU nodes, motivating cheaper atmospheric surrogates for long-duration coupled simulations \cite{chapman2025camulator}. The challenge is to reduce atmospheric cost without sacrificing the ocean--atmosphere feedbacks needed to represent climate variability.

A longstanding alternative is to prescribe atmospheric forcing rather than simulate atmospheric dynamics. CORE-I/II and the CMIP6 Ocean Model Intercomparison Project (OMIP), prescribe atmospheric fields from observations or reanalyses, while surface fluxes are computed from bulk formulae using the prescribed atmosphere and evolving ocean--ice state \cite{griffies2009coordinated,griffies2016omip}. JRA55-do provides the interannually varying atmospheric forcing for OMIP, with fluxes calculated using the Large--Yeager formulation \cite{tsujino2018jra,large2009global}. In CESM, the Data Atmosphere (DATM) supplies prescribed forcing to components such as POP2 without a responsive atmosphere \cite{danabasoglu2020community,cdeps2024software}. Thus, surface fluxes respond to the evolving ocean state, but the prescribed atmosphere does not respond to ocean anomalies, limiting atmospheric feedbacks \cite{griffies2009coordinated}. Intermediate-complexity models retain atmospheric feedbacks through simplified atmospheric components. Zebiak--Cane coupled a simple atmosphere to a reduced-gravity ocean, while Neelin coupled a simplified tropical-Pacific ocean GCM to a steady-state atmosphere \cite{zebiak1987model,neelin1990hybrid}. Later, statistical hybrid models used SVD to relate SST anomalies to wind-stress anomalies, followed by U-Net models that learned this mapping directly \cite{zhang2015hybrid,du2024u}. These approaches retain key ocean--atmosphere feedbacks at lower cost, but remain largely regional or intermediate-complexity. 

Recent global machine-learning emulators offer much larger computational gains. CAMulator achieves a reported $350\times$ speedup over CAM6, while ACE2 remains stable in a 1000-year integration \cite{chapman2025camulator,watt2025ace2}. However, these models are large ($450$--$750$M parameters) and require autoregressive atmospheric-state rollouts. They also lack interactive sea ice. When coupled to prognostic oceans, additional stability challenges have been reported: ACE2-SOM develops unphysical transients under non-equilibrium CO$_2$ forcing, while ACE2-NEMO exhibits sea-ice-volume and sea-surface-height drift that motivated replacing learned ice fluxes with CORE bulk formulae \cite{clark2025ace2,antonio2026ace2}. SamudrACE similarly remains computationally intensive and exhibits weaker ENSO variability \cite{duncan2026samudrace}. Conversely, fully learned coupled systems such as Ola and DLESyM, as well as learned ocean components such as Samudra, demonstrate impressive long-term speed and stability, but replace the physics-based ocean with a learned representation as well. \cite{wang2024coupled,cresswell2025deep,dheeshjith2025samudra}.

A central challenge is to replace atmospheric dynamics while retaining surface forcing that responds to the evolving ocean--ice state. FASST addresses this by coupling a learned atmosphere to the physics-based POP2 ocean \cite{smith2010parallel,danabasoglu2020community} and a lightweight autoregressive sea-ice emulator. Its three main contributions are: (1) a \textbf{stateless atmospheric formulation} that predicts global surface fluxes directly from the current ocean--ice state without propagating an atmospheric state; (2) a \textbf{compact and efficient implementation} with $7.7$M parameters, requiring $10.4$~ms of pure network inference per $6$-hour step on a single NVIDIA A100 versus $190$~ms for CAMulator's atmospheric inference, yielding an approximately $18\times$ speedup; and (3) \textbf{long-term coupled stability}, demonstrated by a $\TotalSimulationLength$-year free-running integration without SST restoring, with global-mean SST equilibrating within $\SpinupYear$ years and remaining close to the CESM$2$ pre-industrial control climatology.

\section{Methodology}
\label{sec:methodology}

FASST couples a diagnostic atmospheric emulator and an autoregressive sea-ice emulator to the physics-based POP2 ocean model at $1^\circ$ resolution. The two emulators are trained independently with no direct coupling between them and are integrated together only during the coupled simulation through their interaction with the ocean and sea-ice states. Both emulators use the U-Net architecture \cite{ronneberger2015u}, originally developed for biomedical image segmentation, which we adapt here for global climate fields. We replace the original padding with circular padding in longitude and replicated padding in latitude to preserve periodicity and spatial resolution. Strided convolutions perform downsampling, while bilinear interpolation followed by convolution performs upsampling. Each stage uses convolution, BatchNorm, and nonlinear activation, with a residual connection from the initial projection to the final layer. A final $1\times1$ convolution produces the continuous physical fields required for climate modeling.

\textbf{Atmospheric emulator:} The emulator maps six inputs: sea surface temperature (SST), sea-ice fraction, solar insolation, 24-hour-lagged SST and sea-ice fraction, and SST tendency, defined as $(\mathrm{SST}_t-\mathrm{SST}_{t-24\mathrm{h}})/(86400~\mathrm{s})$, to $13$ atmospheric fields spanning air-sea fluxes and near-surface atmospheric variables, including wind stress, sensible and latent heat fluxes, evaporative freshwater flux, radiative fluxes, precipitation, near-surface winds, air temperature, humidity, and surface pressure. The lagged SST and sea-ice fraction are provided as input features and are not predicted by the emulator, so the atmospheric emulator is not autoregressive. The emulator is trained on 6-hourly output from the Community Earth System Model version 2.1.5 (CESM2.1.5), generated using the CREDIT pipeline \cite{schreck2025community}. The data are remapped from the CAM grid ($192\times288$) to the POP2 gx1v7 grid ($384\times320$). We use the same chronological 34/1/1-year train/validation/test split as CAMulator \cite{chapman2025camulator}. Training for 80 epochs takes $7$ hours on an NVIDIA A100-SXM4-80GB. We report the coefficient of determination $R^2$ (see Appendix~\ref{app:evaluation}), which measures the fraction of target variance explained relative to the pooled test-set mean. $R^2$ is computed by pooling all held-out grid cells and $6$-hourly test timesteps for each variable. Most fields achieve $R^2>0.9$, with wind stress at $R^2=0.63$--$0.73$ and precipitation at $R^2=0.48$.

\textbf{Sea-ice emulator:} autoregressively predicts monthly ice area fraction, thickness, velocity, and ice--ocean stresses from SST, solar insolation, and the previous month's ice state. It is trained on monthly-mean CESM2-LE2-1231.002 output \cite{rodgers2021ubiquity}, as these prognostic ice fields are unavailable in the CESM$2$ CREDIT data. We use a masked MSE loss and AdamW, with an $80/10/10$ chronological train/validation/test split. Training for 80 epochs with batch size 8 takes about one hour on an NVIDIA A100-SXM4-40GB. The emulator achieves $R^2=0.97$ for ice area fraction and thickness and $R^2=0.53$--$0.65$ for velocity and ice--ocean stress. $R^2$ is evaluated on the monthly-mean fields. Training details and variable-wise scores for both emulators are provided in Appendices~\ref{app:details} and~\ref{app:evaluation}.

\textbf{Coupling:} POP2 provides the prognostic ocean dynamics, while the emulators represent the atmospheric and sea-ice processes needed to close surface exchanges. POP2 runs in ocean-only configuration, with the standard CPL6 exchange (the CESM coupler framework) replaced by direct coupling to the emulators. Every $6$ simulated hours, POP2 supplies SST to the atmospheric emulator, which predicts surface fluxes for the next interval. The sea-ice emulator predicts monthly ice fields from SST and the previous ice state; these are linearly interpolated to modulate the atmospheric fluxes between updates, allowing ocean SST to feed back through the atmosphere and sea ice onto POP2.

\section{Results}
\label{sec:results}
\begin{figure}[t]
  \centering
  \includegraphics[width=\linewidth]{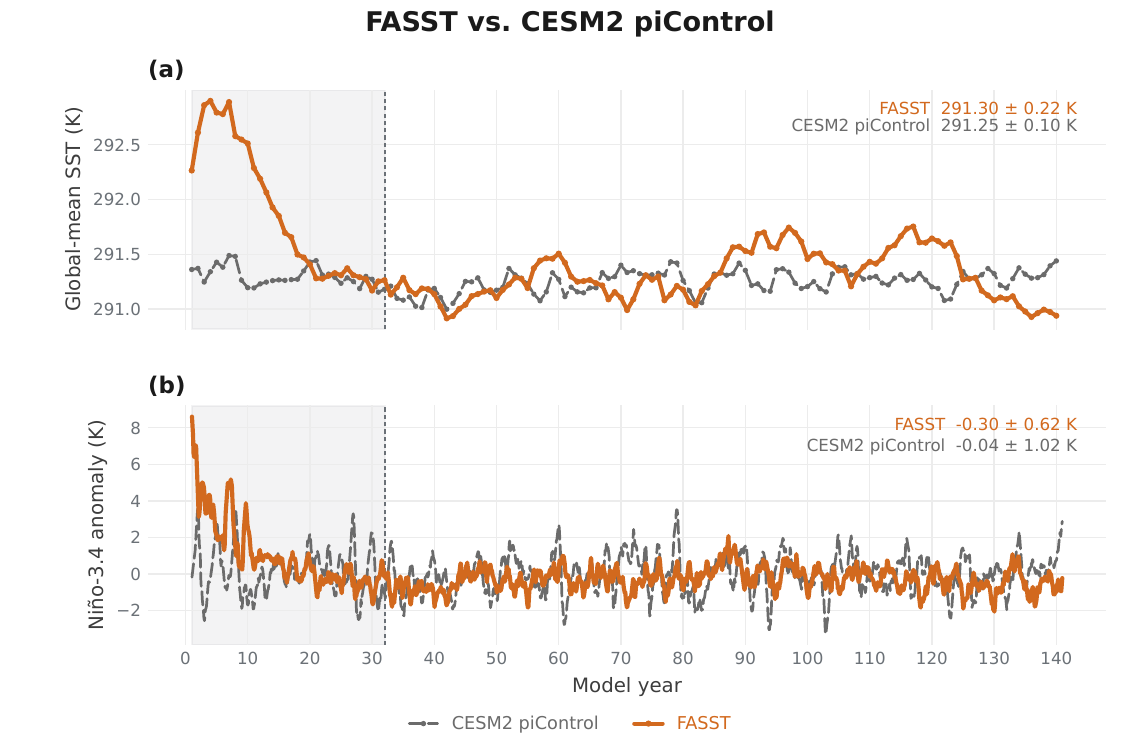}
  \caption{Long-term behavior of FASST (orange, solid) and CESM$2$ piControl (grey, dashed) over $\TotalSimulationLength$ years. The shaded region denotes the initial spin-up, and the vertical dashed line marks its end at year~$\SpinupYear$. (a) Global-mean SST. Following spin-up, FASST has a mean of $\FASSTSSTmean \pm \FASSTSSTstd~\mathrm{K}$ compared with $\CESMSSTmean \pm \CESMSSTstd~\mathrm{K}$ for piControl. (b) Ni\~no-$3.4$ SST anomaly. FASST captures interannual variability but with lower amplitude, with standard deviations of $\FASSTNinoStd$ and $\CESMNinoStd~\mathrm{K}$, respectively.}
  \label{fig:sst_nino}
\end{figure}

We evaluated FASST using an unforced cold-start simulation of \(\TotalSimulationLength\) years and compared its climate variability and mean state with CESM$2$ piControl as a reference for internally generated variability \cite{danabasoglu2020community}. As shown in Figure~\ref{fig:sst_nino}a, FASST undergoes an initial warm adjustment, reaching nearly $293~\mathrm{K}$ before settling into the piControl range. We define the end of this transient as the first year after which the trailing $\SpinupWindow$-year rolling trend in global-mean SST remains below $\SpinupThreshold~\mathrm{K\,yr^{-1}}$, which occurs after $\SpinupYear$ years. Following this adjustment, global-mean SST is $\FASSTSSTmean \pm \FASSTSSTstd~\mathrm{K}$, close to $\CESMSSTmean \pm \CESMSSTstd~\mathrm{K}$ in piControl, indicating a stable global-mean climate without SST restoring. The tropical Pacific exhibits weaker Niño-$3.4$ variability (Figure~\ref{fig:sst_nino}b), with a standard deviation of $\FASSTNinoStd~\mathrm{K}$ versus $\CESMNinoStd~\mathrm{K}$ in piControl, consistent with weaker tropical variability reported for other coupled ML atmosphere emulators, including ACE2-NEMO and CAMulator \cite{chapman2025camulator,antonio2026ace2}.

\textbf{Model Speed Comparison:} Across three runs, FASST achieves $160.9$ SYPD (simulation years per day) for atmosphere-only inference on a single A100. The pure network inference time is $10.4$~ms per $6$-hour step, compared with $190$~ms for CAMulator's atmospheric inference \cite{chapman2025camulator, chapman2026talk}, corresponding to an $18\times$ speedup for the atmospheric component. In the full coupled configuration, using 2 CPU nodes and 1 A100, FASST achieves $35$ SYPD, compared with $42$ SYPD reported for CAMulator on the same hardware configuration \cite{chapman2026talk}. We hypothesize that the lower end-to-end throughput is partly attributable to POP2 handling, further system-level optimization may improve overall throughput.

\begin{figure}[t]
  \centering
  \includegraphics[width=0.8\linewidth]{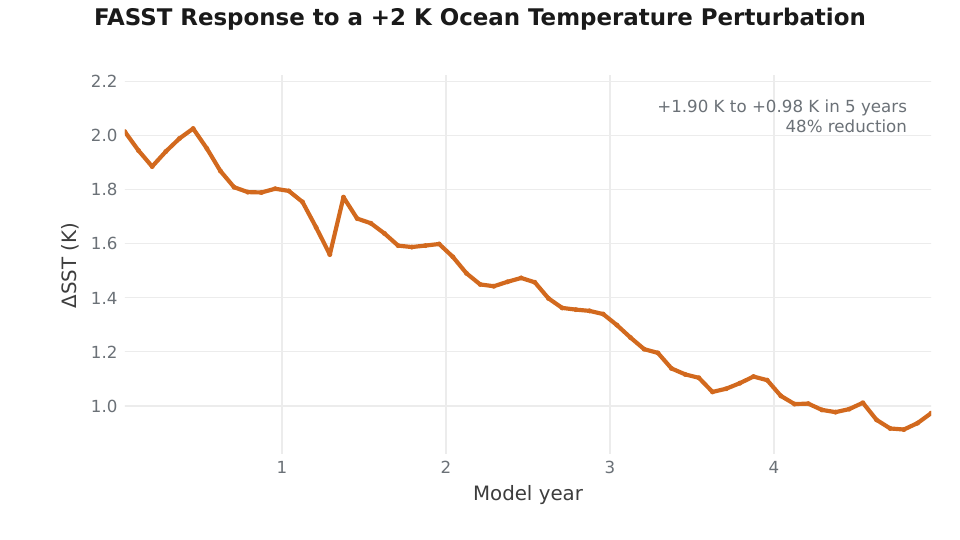}
  \caption{Annual-mean global SST anomaly following a $+2~\mathrm{K}$ perturbation to the
  top $100~\mathrm{m}$ of the initial ocean state, relative to an identical unperturbed
  control. The anomaly decays from $+1.90~\mathrm{K}$ at model year 0 to $+0.98~\mathrm{K}$
  at model year $4$, a $48$\% reduction, with no runaway amplification.}
  \label{fig:temp_pert}
\end{figure}

\textbf{Temperature Perturbation:} To test whether FASST damps an imposed ocean heat anomaly, we perturbed the initial POP2 ocean state by $+2~\mathrm{K}$ over the upper $100~\mathrm{m}$ and integrated for five model years with all other settings unchanged. The anomaly decayed from $+1.90~\mathrm{K}$ at year~$0$ to $+0.98~\mathrm{K}$ at year~$4$, corresponding to a $48\%$ reduction with no runaway amplification observed over the five-year integration (Figure~\ref{fig:temp_pert}). This test probes FASST's coupled response on a shorter timescale than the multi-decadal adjustment from a cold start.

\section{Conclusion}
FASST demonstrates a stateless approach to learned atmospheric coupling, in which surface fluxes are predicted directly from the current ocean--ice state and prescribed insolation. Coupled to POP2 and a learned sea-ice component, FASST has been demonstrated to run stably for $\TotalSimulationLength$ years without SST restoring, maintains a global-mean SST close to CESM$2$ piControl, and reproduces interannual Ni\~no-$3.4$ variability, although with weaker amplitude. Regional SST biases remain, most notably a colder tropical Pacific (Figure~\ref{fig:sst_climatology}). A perturbation experiment further shows that an imposed ocean heat anomaly decays without runaway amplification. Together, these results show that a stateless learned atmosphere can support stable, long-duration coupled climate simulations. Future work will evaluate FASST beyond SST, including ocean heat content, salinity, the Atlantic Meridional Overturning Circulation (AMOC), and the spatial structure of these fields. Moreover, we plan to include CO$_2$ as an input to the model to enable more diverse climate change experiments.

\section{Impact statement}
FASST addresses a key computational bottleneck in long-duration climate experiments. Its compact $7.7$M-parameter, non-autoregressive atmosphere trains in seven hours on a single A100, reducing the resources needed for model development. This makes longer simulations, larger ensembles, and uncertainty studies of ocean heat uptake, variability, and sea-ice processes more accessible. Because the atmosphere is cheap to fine-tune or retrain, researchers can run controlled experiments by modifying it and coupling it to the ocean. Hence, we hope this simple, stateless, yet realistic atmosphere model will be a useful tool for probing coupled atmosphere--ocean interactions.

\bibliographystyle{unsrt}
\bibliography{references}

\appendix
\section{Model and training details}
\label{app:details}

\subsection{Data and code availability}
\label{app:datasets}

The atmospheric emulator is trained on 6-hourly output from Community Earth System Model version 2.1.5, produced for the CREDIT pipeline (\url{https://www.mmm.ucar.edu/opensky-publications/articles:43854/community-research-earth-digital-intelligence-twin-a-scalable-framework-for-ai-driven-earth-system-modeling}) \citep{schreck2025community}. The sea-ice emulator is instead trained on monthly output from a CESM2 Large Ensemble member (1231.002) (\url{https://www.cesm.ucar.edu/community-projects/lens2}) \citep{rodgers2021ubiquity}, since the prognostic sea-ice fields it requires are not available in the CESM2 archive used for the atmospheric emulator. As an unforced baseline, we use the CESM2 piControl simulation (\url{https://data.ucar.edu/dataset/cmip6-cesm2-picontrol-experiment-with-cam6-interactive-land-clm5-coupled-ocean-pop2-with-biogeo}) for comparison against the FASST integration \citep{danabasoglu2020community}. The code used to develop and evaluate FASST is available at \url{https://github.com/pratham-aggr/couple_modeling.git}

\subsection{Training}
\label{app:training}

Both emulators are trained using a masked mean-squared error (MSE) over ocean
grid points,
\[
\mathrm{MSE}
=
\frac{1}{|\Omega|}
\sum_{i\in\Omega}
\left(y_i-\hat{y}_i\right)^2,
\]
where $y_i$ and $\hat{y}_i$ are the reference and predicted values, respectively,
and $\Omega$ denotes the set of points in the ocean grid.All input and target variables are standardized using the training-set mean and standard deviation and transformed back to physical units for evaluation.

\begin{table}[H]
\caption{Training hyperparameters and computational cost for the atmospheric and sea-ice emulators.}
\label{tab:hyperparameters}
\centering
\renewcommand{\arraystretch}{0.9}
\setlength{\tabcolsep}{4pt}
\begin{tabular}{lcc}
\toprule
    Hyperparameter & Atmospheric & Sea-ice \\
\midrule
    Loss & MSE & MSE \\
    Optimizer & AdamW & AdamW \\
    Learning rate & $3\times10^{-4}$ & $2\times10^{-4}$ \\
    Weight decay & $10^{-4}$ & $10^{-4}$ \\
    LR schedule & Cosine & Cosine \\
    Epochs & 80 & 80 \\
    Batch size & 8 & 8 \\
    Base width & 64 & 64 \\
    Dropout & 0.1 & 0.1 \\
    Training time & 7 h & 1 h \\
    Hardware & A100-SXM4-80GB & A100-SXM4-40GB \\
\bottomrule
\end{tabular}
\end{table}

\subsection{Network architecture}
\label{app:architecture}

Both emulators use a five-level U-Net but differ in their inputs, outputs, and coupling: the atmospheric emulator is diagnostic, whereas the sea-ice emulator is autoregressive. Input and output variables are listed in Table~\ref{tab:variables}. The atmospheric emulator has $7,726,861$ trainable parameters, a base width of $64$, and $12$ convolutional and 10 batch-normalization layers.

The network begins with a \(1\times1\) convolution that projects the six input channels to 64 feature channels. The encoder comprises five stride-2 convolutional blocks with channel widths of $512$, $256$, $128$, $128$, and $64$. The first four blocks use \(4\times4\) kernels, while the deepest uses a \(2\times2\) kernel. Each block consists of a convolution, batch normalization, LeakyReLU activation with slope 0.2, and dropout with rate 0.1.

The decoder mirrors the encoder with channel widths of $512$, $256$, $128$, $128$, and $64$. Each block bilinearly upsamples the feature map by a factor of two followed by a stride-$1$ convolution. Decoder convolutions use \(3\times3\) kernels, except at the deepest level, where a \(1\times1\) kernel is used. Each is followed by batch normalization, ReLU activation, and dropout with rate $0.1$. Skip connections concatenate encoder and decoder features at matching resolutions.

The final $64$-channel representation is concatenated with the features from the initial projection layer and passed through a \(1\times1\) convolutional head to produce the $13$ output fields. All spatial convolutions use circular padding in longitude and replication padding in latitude, preserving the periodic zonal boundary of the tripole grid without an artificial seam. Convolutional weights are initialized from \(\mathcal{N}(0,0.02^2)\), batch-normalization scales from \(\mathcal{N}(1,0.02^2)\), and batch-normalization biases to zero.

\begin{table}[H]
  \caption{Input and target variables for the two emulators. The atmosphere emulator
  is diagnostic and operates every $6$ hours; the sea-ice emulator is autoregressive
  and operates monthly.}
  \label{tab:variables}
  \centering
  \small
  \begin{tabular}{llll}
    \toprule
    \multicolumn{4}{l}{\textbf{Atmosphere emulator} ($6$-hourly, diagnostic)} \\
    \midrule
    Role & Variable & Description & Units \\
    \midrule
    Input  & SST            & Sea-surface temperature (from POP2)           & K \\
    Input  & ICEFRAC        & Sea-ice fraction (from sea-ice emulator)      & 0--1 \\
    Input  & SOLIN          & Incoming solar radiation, top of atmosphere   & W\,m$^{-2}$ \\
    Input  & SST\_prev      & Sea-surface temperature, $24$\,h lag            & K \\
    Input  & ICEFRAC\_prev  & Sea-ice fraction, $24$\,h lag                   & 0--1 \\
    Input  & dSST\_dt       & SST tendency                              & K\,s$^{-1}$ \\
    \addlinespace
    Target & TAUX    & Zonal wind stress                    & N\,m$^{-2}$ \\
    Target & TAUY    & Meridional wind stress               & N\,m$^{-2}$ \\
    Target & SHFLX   & Sensible heat flux                   & J\,m$^{-2}$ \\
    Target & LHFLX   & Latent heat flux                     & J\,m$^{-2}$  \\
    Target & QFLX    & Surface freshwater flux              & kg\,m$^{-2}$\,s$^{-1}$ \\
    Target & FSDS\_J & Downwelling shortwave radiation      & J\,m$^{-2}$ \\
    Target & FLDS\_J & Downwelling longwave radiation       & J\,m$^{-2}$ \\
    Target & PRECT   & Total precipitation                  & m \\
    Target & Ubot     & Near-surface zonal wind                  & m\,s$^{-1}$ \\
    Target & Vbot     & Near-surface meridional wind             & m\,s$^{-1}$ \\
    Target & Tbot     & Near-surface air temperature              & K \\
    Target & Qbot     & Near-surface specific humidity            & kg\,kg$^{-1}$ \\
    Target & PS       & Surface pressure                         & Pa \\
    \midrule
    \multicolumn{4}{l}{\textbf{Sea-ice emulator} (monthly, autoregressive)} \\
    \midrule
    Input & SST         & Sea-surface temperature (from POP2)               & K \\
    Input & SOLIN       & Top-of-atmosphere insolation (analytic, date+lat) & W\,m$^{-2}$ \\
    Input & aice\_prev  & Sea-ice area fraction, previous month             & 0--1 \\
    Input & hi\_prev    & Sea-ice thickness, previous month                 & m \\
    Input & uvel\_prev  & Sea-ice zonal velocity, previous month            & m\,s$^{-1}$ \\
    Input & vvel\_prev  & Sea-ice meridional velocity, previous month       & m\,s$^{-1}$ \\
    \addlinespace
    Target & aice    & Sea-ice area fraction          & 0--1 \\
    Target        & hi      & Sea-ice thickness              & m \\
    Target       & uvel    & Sea-ice zonal velocity         & m\,s$^{-1}$ \\
    Target       & vvel    & Sea-ice meridional velocity    & m\,s$^{-1}$ \\
    Target      & strocnx & Zonal ice-to-ocean stress      & N\,m$^{-2}$ \\
    Target     & strocny & Meridional ice-to-ocean stress & N\,m$^{-2}$ \\
    \bottomrule
  \end{tabular}
\end{table}

\section{Evaluation}
\label{app:evaluation}

We evaluate emulator performance on held-out test data using the coefficient of
determination ($R^2$) and root mean squared error (RMSE), computed in physical
(de-normalized) units. $R^2$ measures how well the emulator reproduces spatial
and temporal variability, while RMSE measures the magnitude of prediction
errors in the variable's native units. The two metrics are complementary:
$R^2$ is scale-independent, whereas RMSE reflects the absolute physical error.

The metrics are defined as
\begin{equation}
R^2 = 1-\frac{\sum_i(y_i-\hat{y}_i)^2}
{\sum_i(y_i-\bar{y})^2},
\qquad
\mathrm{RMSE}=
\sqrt{\frac{1}{N}\sum_i(y_i-\hat{y}_i)^2},
\end{equation}
where $y_i$ and $\hat{y}_i$ are the reference and predicted values,
$\bar{y}$ is the test-set reference mean, and $N$ is the number of evaluated
points. For each variable, all unmasked valid grid cells and timesteps are pooled,
with no spatial area weighting.

RMSE is reported in each variable's native units. SHFLX and LHFLX
are evaluated as energy accumulated over the $6$-hour coupling interval,
consistent with the FSDS\_J and FLDS\_J targets. For visualization, the
accumulated energy fluxes are converted to mean fluxes in
$\mathrm{W\,m^{-2}}$ by dividing by the $6$-hour coupling interval.
Table~\ref{tab:results} reports skill for both emulators.

\begin{figure}[t]
    \centering
    \includegraphics[width=\linewidth]{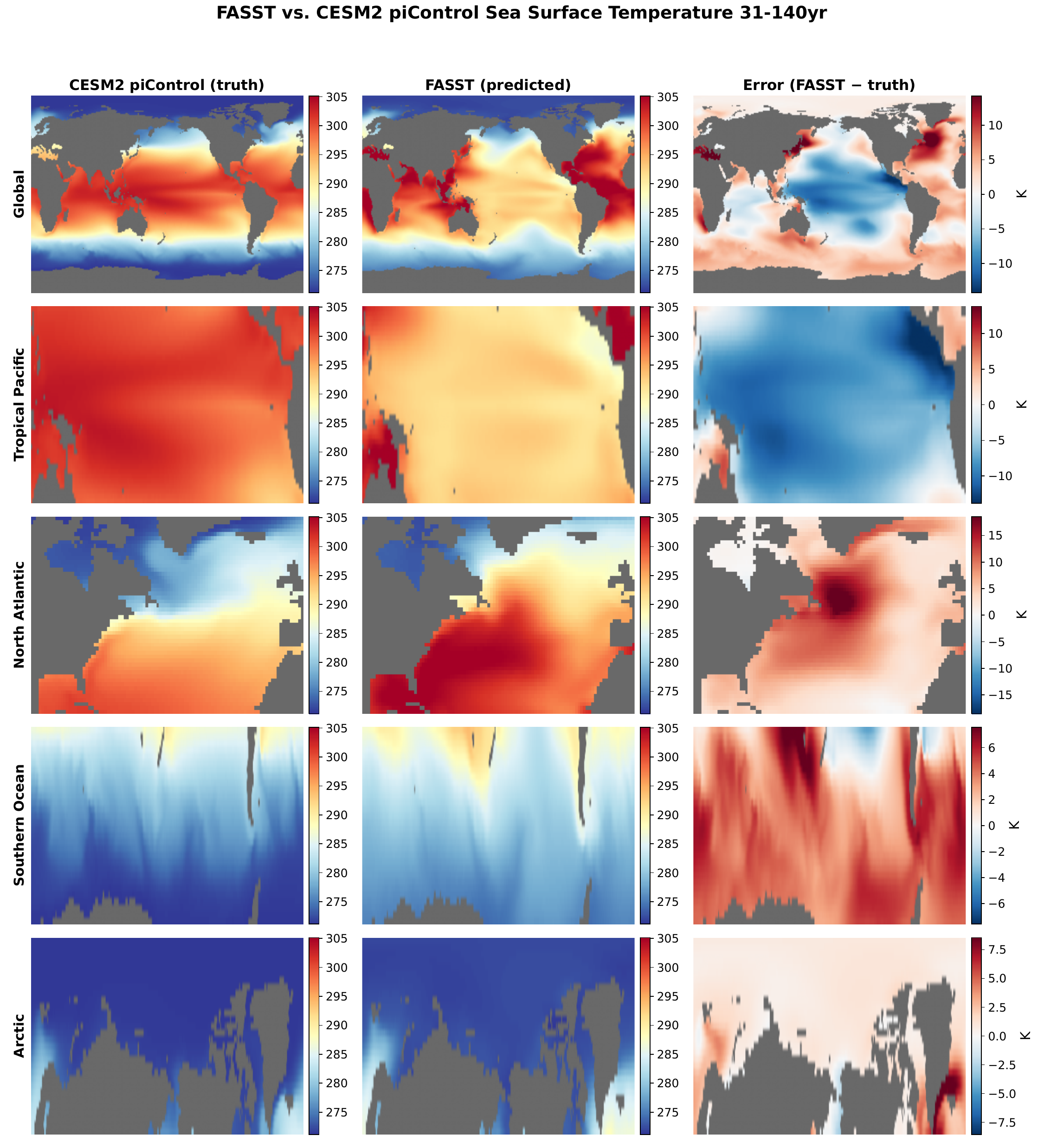}
    \caption{FASST versus CESM$2$ piControl sea-surface temperature climatology.
    Columns show CESM$2$ piControl (truth), FASST (predicted), and the corresponding
    error (FASST minus truth). Rows show the global domain, tropical Pacific,
    North Atlantic, Southern Ocean, and Arctic. Climatologies are computed over
    the post-transient period (model years 31--\TotalSimulationLength) using elapsed-year-matched
    intervals.}
    \label{fig:sst_climatology}
\end{figure}

Figure~\ref{fig:sst_climatology} compares the post-transient SST climatology of
FASST with CESM$2$ piControl over matched elapsed years (31--\TotalSimulationLength). FASST captures the
broad large-scale spatial structure of SST, while regional biases remain,
most notably a colder tropical Pacific, as well as differences in the North
Atlantic, Southern Ocean, and Arctic.

\begin{table}[H]
  \caption{Test-set performance of the atmospheric emulator against CREDIT
data and the sea-ice emulator against CESM2 Large Ensemble member
1231.002.}
  \label{tab:results}
  \centering
  \begin{tabular}{lcc}
    \toprule
    \multicolumn{3}{l}{\textbf{Atmospheric emulator}} \\
    \midrule
    Variable & $R^2$ & RMSE \\
    \midrule
    PS             & 0.9996 & $5.9646\times10^{2}$  \\
    Tbot           & 0.9994 & 1.8736                \\
    FLDS\_J & 0.9840 & $3.2478\times10^{5}$ \\
    Qbot           & 0.9678 & $8.6057\times10^{-4}$ \\
    QFLX    & 0.9243 & $1.6252\times10^{-4}$ \\
    FSDS\_J & 0.9316 & $1.0778\times10^{6}$ \\
    LHFLX  & 0.9186 & $4.0856\times10^{5}$ \\
    SHFLX  & 0.8655 & $3.3691\times10^{5}$ \\
    Ubot           & 0.8042 & 1.8729                \\
    Vbot           & 0.7179 & 1.8921                \\
    TAUX    & 0.7338 & $5.6026\times10^{-2}$ \\
    TAUY    & 0.6275 & $5.2846\times10^{-2}$ \\
    PRECT   & 0.4751 & $1.2382\times10^{-3}$ \\
    \midrule
    \multicolumn{3}{l}{\textbf{Sea-ice emulator}} \\
    \midrule
    Variable & $R^2$ & RMSE \\
    \midrule
    aice     & 0.974 & 0.0341  \\
    hi       & 0.974 & 0.0424  \\
    uvel     & 0.653 & 0.0127  \\
    strocnx  & 0.576 & 0.0075  \\
    strocny  & 0.570 & 0.00562 \\
    vvel     & 0.531 & 0.00966 \\
    \bottomrule
  \end{tabular}
\end{table}

\begin{figure}[H]
  \centering
  \includegraphics[width=\linewidth]{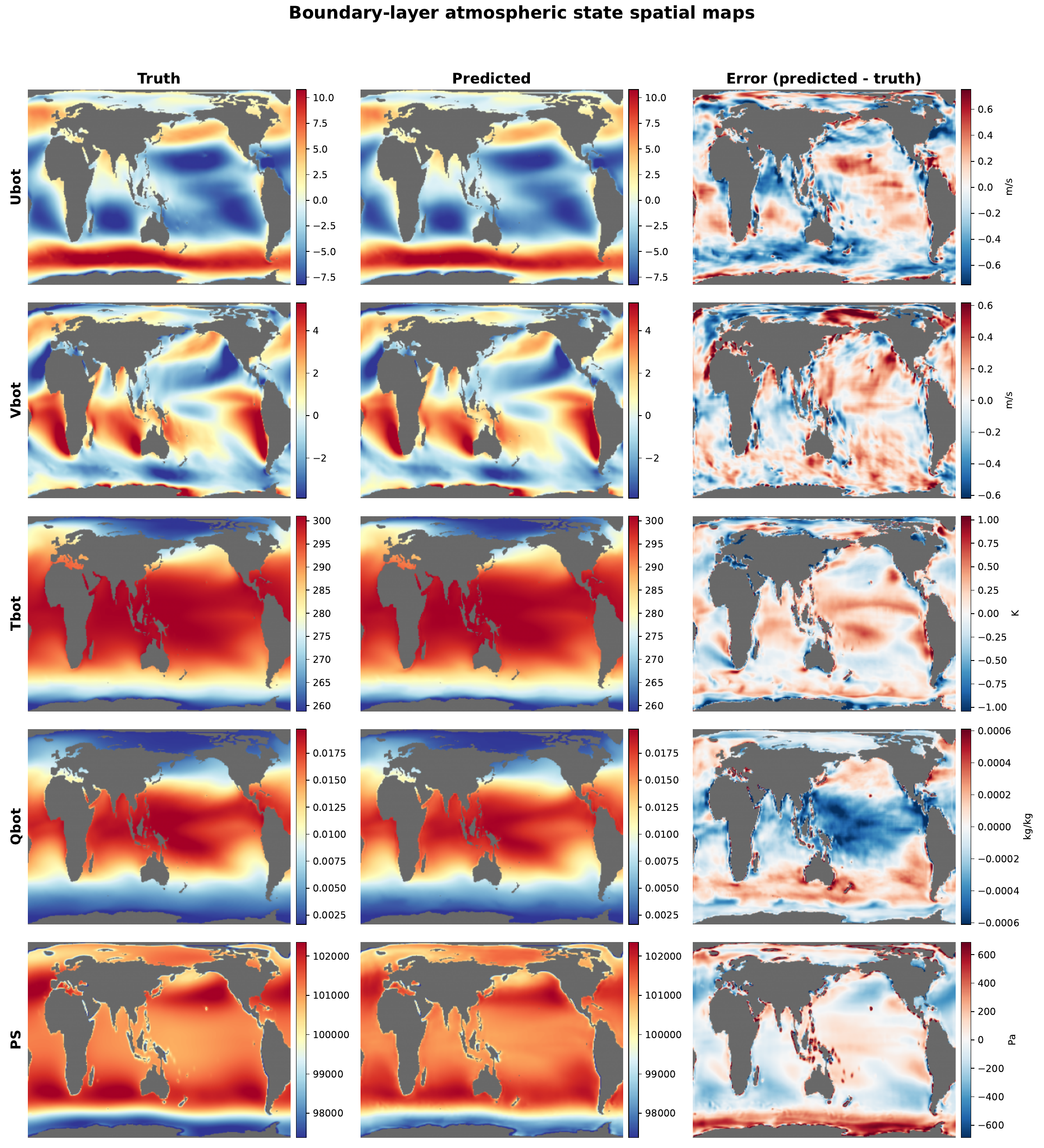}
  \caption{Test-period spatial maps of the atmospheric emulator's
  boundary-layer state outputs (Ubot, Vbot, Tbot, Qbot, PS; $R^2=0.72$--$1.00$):
  reference (left), FASST prediction (middle), and error (right), averaged
  over the held-out test period.}
  \label{fig:atmstate_maps}
\end{figure}

\begin{figure}[H]
  \centering
  \includegraphics[width=\linewidth]{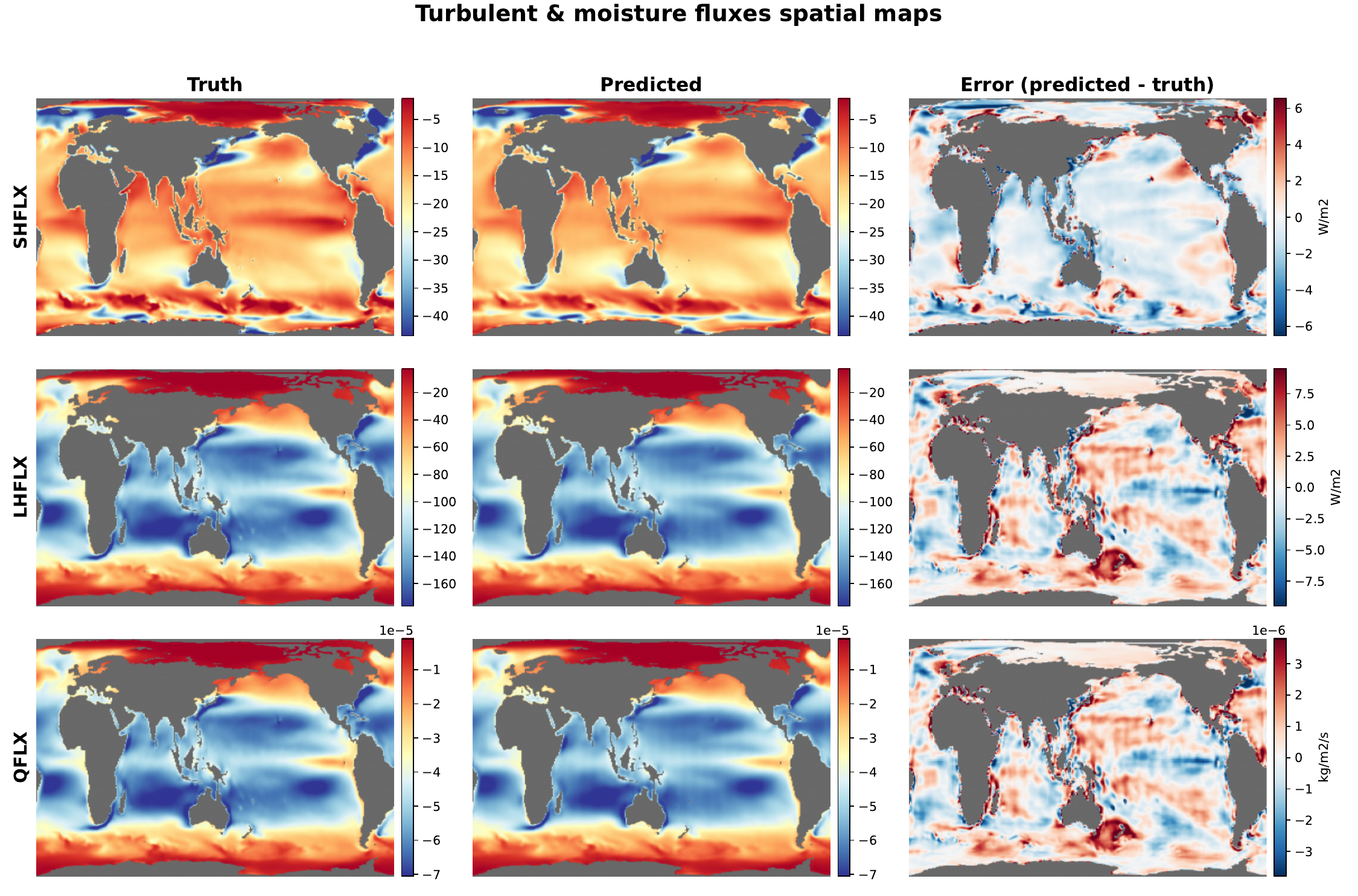}
  \caption{Test-period spatial maps of the atmospheric emulator's turbulent
  and moisture flux outputs (SHFLX, LHFLX, QFLX; $R^2=0.87$--$0.92$):
  reference (left), FASST prediction (middle), and error (right), averaged
  over the held-out test period.}
  \label{fig:flux_maps}
\end{figure}

\begin{figure}[H]
  \centering
  \includegraphics[width=\linewidth]{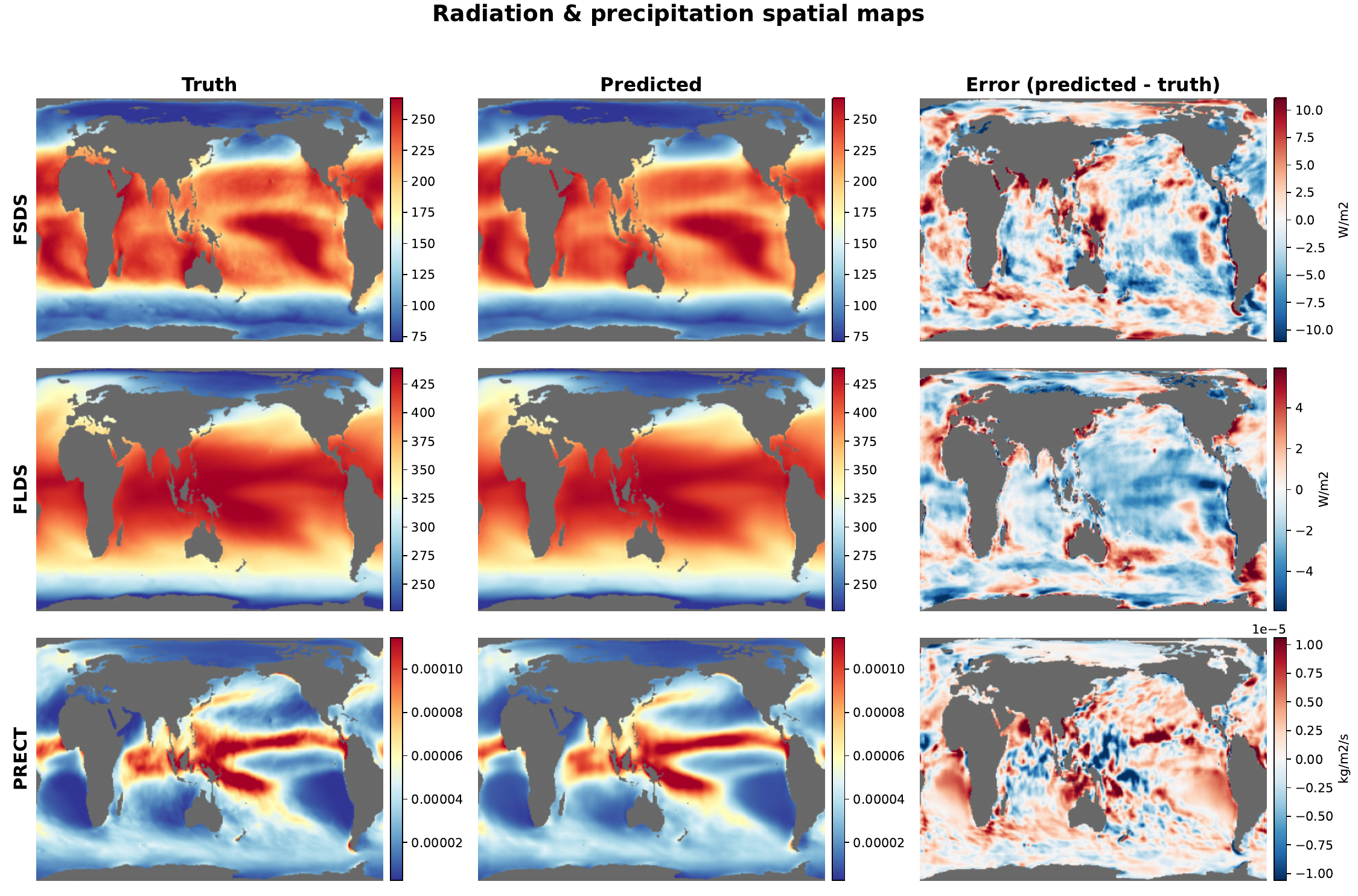}
  \caption{Test-period spatial maps of the atmospheric emulator's radiation
  and precipitation outputs (FSDS, FLDS, PRECT): reference (left), FASST
  prediction (middle), and error (right), averaged over the held-out test
  period. Radiative fluxes are well reproduced ($R^2 \geq 0.93$); PRECT is the
  weakest-performing variable in the atmospheric emulator ($R^2=0.48$).}
  \label{fig:radiation_maps}
\end{figure}

\begin{figure}[H]
  \centering
  \includegraphics[width=\linewidth]{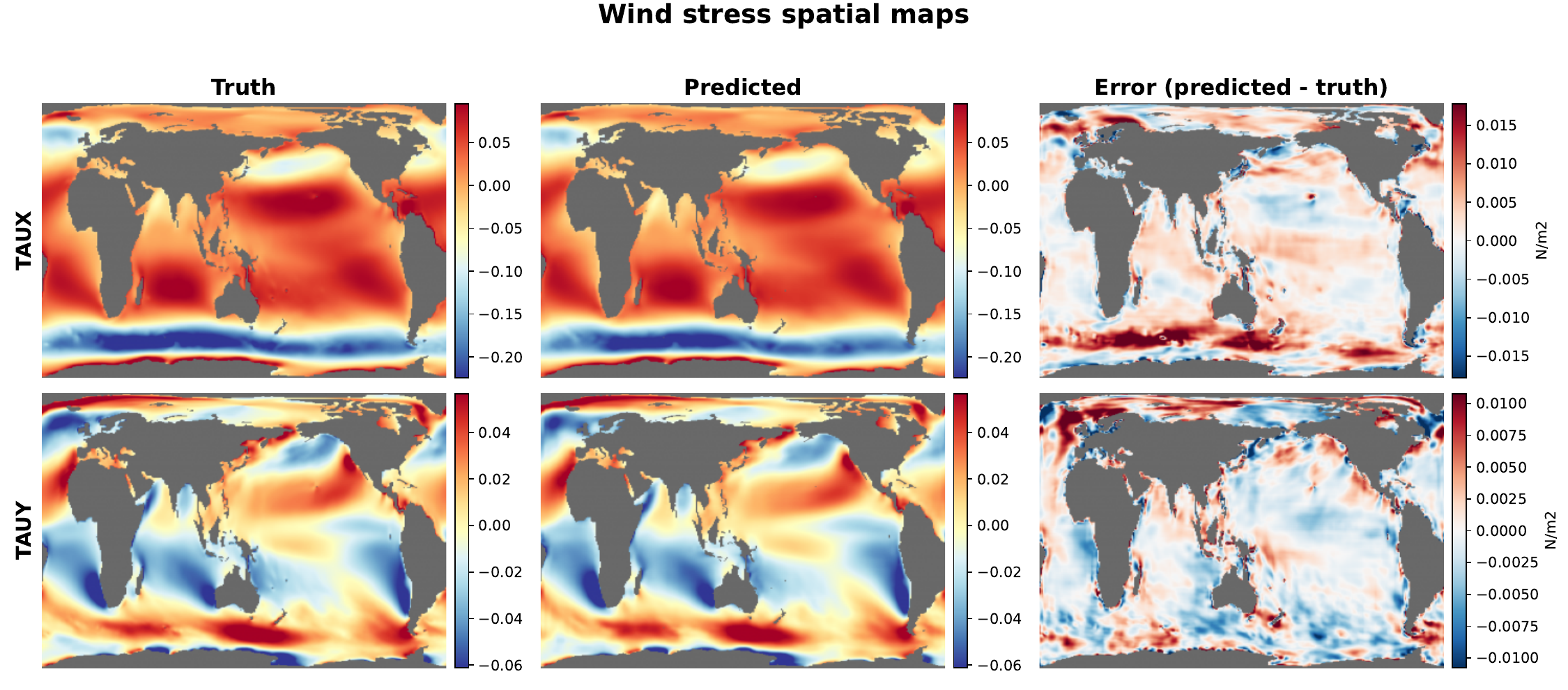}
  \caption{Test-period spatial maps of the atmospheric emulator's wind-stress
  outputs (TAUX, TAUY; $R^2=0.63$--$0.73$): reference (left), FASST
  prediction (middle), and error (right), averaged over the held-out test
  period. These are the weakest-performing forced fields, of interest given
  their direct role in ocean momentum forcing.}
  \label{fig:wind_maps}
\end{figure}

\end{document}